# CONDITIONING AND FUTURE PLANS FOR A MULTI-PURPOSE 805 MHZ PILLBOX CAVITY FOR MUON ACCELERATION*

G. Kazakevich[#], G. Flanagan, R. P. Johnson, M. Neubauer, R. Sah, A. Dudas, F. Mahrhauser.
Muons, Inc., Batavia, IL 60510

A. Moretti, M. Popovic, K. Yonehara, G. Romanov, Fermilab, Batavia, IL 60510

Y. Torun, IIT, Chicago, IL 60616

S. Kurennoy, LANL, Los Alamos, NM 87545

*Abstract*

An 805 MHz RF pillbox cavity has been designed and constructed to investigate potential muon beam acceleration and cooling techniques for a Muon Collider or Neutrino Factory. The cavity can operate in vacuum or under pressure up to 100 atmospheres, at room temperature or in a liquid nitrogen bath at 77K. The cavity has been designed for easy assembly and disassembly utilizing a bolted construction with aluminium seals. To perform vacuum and high-pressure breakdown studies of materials and geometries most suitable for the collider or factory, the surfaces of the end walls of the cavity can be replaced with different materials such as copper, aluminium, beryllium, or molybdenum, and with different geometries such as shaped windows or grid structures. The cavity has been designed to fit inside the 5-Tesla solenoid in the MuCool Test Area (MTA) at Fermilab. In this paper we present the vacuum conditioning results with and without an external magnet field. Additionally, we discuss the future plans for the cavity.

## INTRODUCTION

Ionization cooling, where all momentum components are degraded by an energy absorbing material and only the longitudinal momentum is restored by RF cavities, provides a means to quickly reduce transverse beam sizes. However, the beam energy spread cannot be reduced by this method unless the longitudinal emittance can be transformed or exchanged into the transverse emittance. One scheme to achieve emittance exchange is to pass a beam through a bending magnet to introduce dispersion; the beam can then be made incident to a wedge shaped absorber. The higher momentum particles pass through more of the absorber material than the low momentum particles and thus suffer larger ionization energy losses. Much work has been done on a second cooling scheme, one in which a continuous absorber such as gaseous $H_2$ is used in a Helical Cooling Channel (HCC) [1]. In the HCC higher momentum corresponds to a longer path length. The path length dependence means high momentum particles must pass through more absorber and therefore experience a larger ionization energy loss compared to low momentum particles. The theory of this helical channel has been described elsewhere [2]. The use of high pressure gas not only serves as an ideal absorber, it helps in the suppression of RF breakdown, and if one can design the appropriate thermal barrier between the RF cavity and magnet it will also function as part of the cavity cooling system.

An HCC consisting of a pressurized gas absorber imbedded in a magnetic channel that provides solenoid, helical dipole and helical quadrupole fields has shown considerable promise in providing six-dimensional cooling for muon beams. The energy lost by muons traversing the gas absorber needs to be replaced by inserting RF cavities into the lattice. Replacing the substantial muon energy losses using RF cavities with reasonable gradients will require a significant fraction of the channel length be devoted to RF. However, to provide the maximum phase space cooling and minimal muon losses, the helical channel should have a short period and length.

Demonstrating the technology of such a cooling channel would represent enormous progress toward the next energy frontier machine. The multipurpose 805 MHz cavity described here will facilitate the understanding of how to build a cooling channel. Additionally it is conceptually compatible with another Muons, Inc. proposal that aims to design and build a 10 T, 805 MHz segment of a helical cooling channel which builds on previous work by Muons, Inc. [3].

The cavity can also serve as a test resonator, [4], for a large- acceptance high-gradient linac for acceleration of low energy muons and pions in a strong solenoidal magnetic field. Such a linac was proposed at LANL. The acceleration starts immediately after collection of pions from a target by solenoidal magnets and brings muons to a kinetic energy of about 200 MeV over a distance of the order of 10 m. At this energy, both an ionization cooling of the muon beam and its further acceleration in a superconducting linac become feasible. The required large longitudinal and transverse acceptances can be achieved in a normal-conducting linac consisting of independently fed $TM_{010}$ mode RF cavities with wide apertures closed with thin Be metal windows or grids.

___________________________________________

*Work supported under U.S. DOE Grant DE-FG-08ER86352.

# grigory@muonsinc.com; gkazakevitch@yahoo.com

The guiding magnetic field is provided by external superconducting solenoids. Due to the low energy of the initial pions and muons, vacuum cavities are preferred, at least in the beginning of the normal-conducting linac to minimize particle energy losses.

## FEATURES OF THE CAVITY MODEL INTENDED FOR COOLING MUONS

The work is an attempt to create a universal design of an accelerating cavity model capable of working with 100 atm of compressed hydrogen or with vacuum. To simplify technology and design, the cavity model has been made from thick elements of copper plated stainless steel SS-316 clamped with bolts, Figure 1, [4]. The sealing identical for the vacuum and for compressed gas is provided by flat pure aluminum gasket rings.

We have developed a technology for manufacturing of the flat aluminum gasket rings and machining of the respective sealing surfaces of the cavity parts required for vacuum sealing. Details of the technology and the cavity preparation can be found in [4].

The pillbox cavity operating in the $TM_{010}$ mode consists of two lids, a cylindrical body and two test plate holders, which allow using test plates as insets to check various materials for operation within a high-gradient RF field. All the elements are plated with 25-37 µm thick copper.

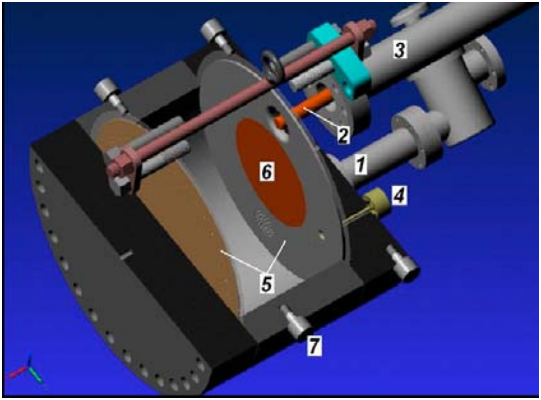

Figure 1: The model of the cavity for cooling of muons, 1-vacuum port, 2-antenna, 3-RF coax feeder, 4- probe, 5- test plateholder, 6-test plate to study materials at high gradients, 7- roller.

Results of the cavity model vacuum test and some parameters of the cavity model are presented in [4].

## CONDITIONING PERFORMANCE

The cavity was conditioned in a few runs in the MTA at the Fermilab. The initial runs were performed without an external magnetic field. The cavity was fed by a 805 MHz klystron pulsed transmitter. The cavity was coupled with the klystron waveguide by matched waveguide-coaxial and coaxial-coaxial adapters having commercial vacuum windows for sealing the cavity vacuum system. Outside the vacuum the adapter was pressurized to ~14 psi of $SF_6$ to avoid breakdowns in the windows. The cavity on the RF bench is shown in Figure 2.

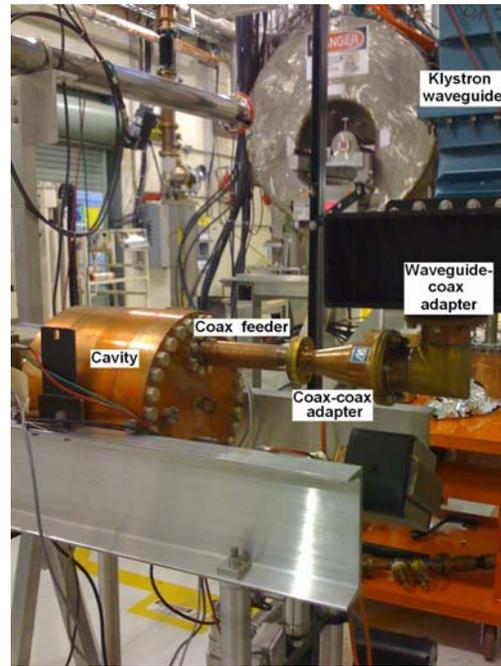

Figure 2: The RF cavity model on the RF bench in the MTA at the Fermilab.

Conditioning without an external magnetic field was initially performed at the RF pulse duration of 20 µs and repetition rate of 15 Hz. In this regime the cavity was conditioned to electric field gradient of 16.4 MV/m. Further conditioning was performed at repetition rate of 2 Hz with the same pulse duration. The field gradient value was measured using calibrated directional couplers in the klystron waveguide system.

Recently the cavity was moved into the MTA superconducting solenoid bore, Figure 3. The cavity axis was aligned parallel to the solenoid axis. A comparison of the operation of the cavity in a magnetic field of 3 T and without the field was done.

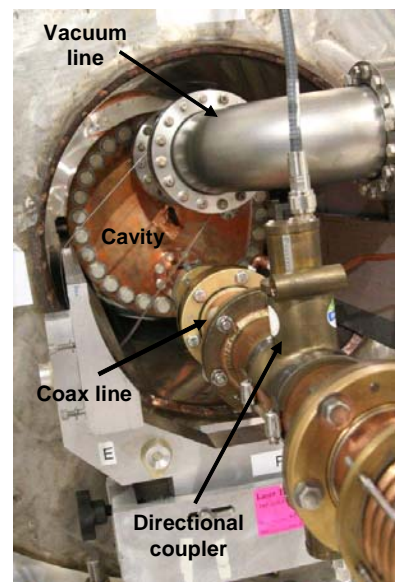

Figure 3: Front view of the cavity model located in bore of the superconducting solenoid in MTA.

In both cases we have reached gradient of electric field of 25 MV/m with or without the magnetic field of 3 T being on. This was studied for several long runs and the same results were achieved. Trying to operate slightly above 25 MV/m caused the reflected power signal to increase by about of 25 to 30 % and have oscillating pattern similar to what is seen in multipactor (MP) or dark current loading of the cavity. The effect went away a soon as we lowered the gradient back to 25 MV/m. The effect occurred immediately with not soft turn on and it turned off with no soft turn off.

This result differs from those achieved for the LBNL cavity, an 805 MHz pillbox. Here a peak electric field of 34 MV/m was obtained without external magnetic field, yet the gradient was limited to 16 MV/m in a 3 T external magnetic field. There are significant geometrical differences between the two cavities that may explain the differing performance limitations observed. The LBNL cavity is 8.1 cm long and utilizes a waveguide input coupler, whereas the Muons, Inc. cavity is ~15 cm long and uses a coaxial capacitive coupling.

Understanding the differences in performance between the LBNL and Muons, Inc. cavities has become a very important step to defining/refining directions for muon acceleration cavity R&D. Now we are using simulation tools to understand the performance differences. The next steps toward understanding the differences are described below.

## FUTURE WORK TOWARD UNDERSTANDING PERFORMANCE LIMITATIONS IN A MAGNETIC FIELD

Simulation work has begun to understand the performance limitations of the Muons, Inc. cavity using 2D FishPact [6] and 3D ACE3P/Track3P [7]. The focus of these studies was to reveal potential multipacting (MP) sites and barriers within the cavity. Sustained secondary electron trajectories (surviving at least 20 RF cycles) have been found with impact energies peaking around 23.6 MV/m, which end on the outer cylinder of the cavity wall (1-point MP, 1$^{st}$ order). The impact energies however are very low (< 20 eV, SEC <1 for copper), i.e. to not induce a sustained MP, Figure 4.

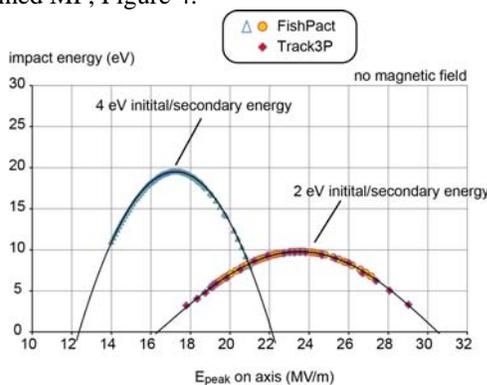

Figure 4: Impact energy vs. electric field for stable 1-point (1$^{st}$ order) resonant trajectories ending on the cavity cylindrical wall without solenoidal field. Once a 3T is applied however, these resonances disappear.

Hereby the impact energy depends also on the initial/secondary energy assumed. We have taken into account 2eV and 4eV, respectively. Moreover, the resonant trajectories completely ceased once a 3T solenoidal field was applied.

A 2-point MP (1$^{st}$ to few higher orders) between the cavity end lids as theoretically possible between parallel plates can exist only at much lower fields that cannot explain the experimental limitations seen at 25 MV/m. Our focus is therefore on field emission loading as a possible operational limitation. A high magnetic solenoidal field can strongly focus electron trajectories on the path between the end lids as verified with both codes mentioned above. Corresponding simulations are still in progress including estimates of field emission induced heating of the end plates. We intend to compare the findings with those obtained in the LBNL cavity. Understanding the differences in performance between the LBNL and Muons, Inc. cavities has become a very important step to defining/refining directions for muon acceleration cavity R&D.

## SUMMARY

A test model of the RF cavity for the HCC tasks has been developed, built and tested with an 805 MHz klystron. Conditioning without external magnetic field demonstrated operation of the cavity with a maximum electric field gradient of 25 MV/m. Conditioning the cavity in the external 3 T solenoidal magnetic field a maximum gradient of electric field of 25 MV/m was reached. Simulations including estimates of field emission induced heating of the end plates are still in progress.